\documentclass{KapProc} 

\normallatexbib

\begin{document}

\articletitle{Complexity at Mesoscale}

\articlesubtitle{}

\author{A.~R. Bishop, K.~\O. Rasmussen, J. R{\"o}der, T. Lookman, and A. Saxena}
\affil{Theoretical Division and Center for Nonlinear Studies, Los Alamos National Laboratory,\\
Los Alamos, New Mexico 87545, USA}
\email{}

\author{Andrea Vanossi}
\affil{Dipartimento di Fisica, Universit\`a di
  Bologna, V.le Berti Pichat 6/2, I-40127, Bologna,
  Italy, and INFM e Dipartimento di
  Fisica, Universit\`a di Modena, Via Campi 213/A, 41100 Modena, Italy}

\author{Panayotis Kevrekidis}
\affil{Program in Applied and Computational Mathematics, Princeton University, Fine Hall,
Washington Road, Princeton, New Jersey, 08544-1000, USA}
\email{}


\begin{abstract} 
Through three examples we illustrate some of the concepts and ingredients required for 
pattern formation at mesoscopic scales. Two examples build on microscopic models where
mesoscopic patterns emerge from homogeneous ground states driven into instability by 
external forcing. In contrast,
the third example builds on a mesoscopic phenomenological Ginzburg-Landau type model of 
solid-solid structural phase transitions. 
Here, mesoscopic textures emerge as a result of competing length scales 
arising from the constraints of 
elastic compatibility.
\end{abstract}

\section{Introduction}
Complex spatial patterns are typical observations in physical systems at mesoscales. Often such mesoscopic complex
spatial patterns are observed in systems driven away from equilibrium or as results of phase transitions. By way of three 
examples, in the present paper we describe some of the ingredients that work together to 
create mesoscopic patterns. We will describe two quite different driven dynamical systems in which 
the uniform ground state becomes unstable as the system is driven away from equilibrium. In both cases we illustrate how 
the instability through nonlinear, and long-range interactions gives rise to dynamically stable mesoscopic patterns 
consisting of locally coherent defect-like entities.  To contrast these two examples we also discuss modeling 
of solid-solid structural 
phase transitions such
as found in martensitic materials 
\cite{R1}. In the latter example our model builds on continuum elasticity in a phenomenological
manner. In this case we will show how temperature (or pressure) driven structural phase transitions result in the 
emergence of mesoscopic textures in strain fields. 

These three examples serve to illustrate several ingredients controlling mesoscopic patterns: driven
versus undriven environments; the importance of discrete lattice scales; the competitions between (often anisotropic)
short and long-range interactions; and the role of constraints as source of long-range interactions.
\section{Pattern formation in a driven nonlinear lattice}
We will start with our simplest example \cite{Andrea}, a driven damped Klein-Gordon 
lattice and consider it in some detail to 
introduce concepts. The explicit model is
\begin{eqnarray}
\ddot x_{n}+\gamma \dot x_n +\omega_0^2 x_n=\Delta_n x_{n}
+\lambda x^3_n+\varepsilon \cos\omega t,
\label{eqs}
\end{eqnarray}
where $\gamma$ is the damping parameter, $\omega_0$ the natural
frequency of the oscillators, $\lambda=\pm 1$ the nonlinearity parameter, and,
finally $\varepsilon$ is the amplitude of the ac-drive at frequency
$\omega$. In one spatial dimension the nearest neighbor coupling is
$\Delta_n x_n=x_{n+1}-2x_n+x_{n-1}$ \cite{2Dana}. The amplitude $A_0$ (and
phase $\delta_0$) of the spatially homogeneous solution $x_n=y=A_0\cos(\omega
t+\delta_0)$ of Eq.(\ref{eqs}) can, within the rotating wave
approximation, be shown to satisfy
\begin{equation}
A_0^2
\left (\gamma^2\omega^2+(\omega^2-\omega_0^2+\frac{3}{4}
\lambda  A_0^2)^2 \right)=\varepsilon^2.
\label{A0}
\end{equation}
Analyzing the stability of the homogeneous solution with respect to
spatial perturbations, we introduce $x_n=y+z_n$ into Eq.(\ref{eqs}).
Assuming periodic boundary conditions, we may expand $z_n$ in its
Fourier components $z_n=\sum_k \exp( i k n) \xi_k(t)$, where the mode
amplitude $\xi_k(t)$ is then governed by
\begin{eqnarray}
\ddot \xi_k+\gamma\dot \xi_k+\omega_k^2\xi_k&=&\frac{3}{2}\lambda A_0^2 \left [
1+\cos(2\omega t +2 \delta_0) \right ] \xi_k,
\label{HILL1} 
\end{eqnarray}
with $\omega_k^2=\omega_0^2+4\sin^2(k/2)$ denoting the linear dispersion
relation.

Finally, the transformation $\xi_k(t)=\zeta_k(\omega
t+\delta_0)\exp(-\frac{\gamma}{2}(\omega t+\delta)) \equiv
\zeta_k(\tau)\exp(-\frac{\gamma}{2}\tau)$ reduces Eq.(\ref{HILL1}) to
a standard Mathieu equation
\begin{eqnarray}
\frac{d^2\zeta_k}{d\tau^2}+a\zeta_k-2q\cos(2\tau)\zeta_k=0,
\label{MATHIEU1}
\end{eqnarray}
where
\begin{equation}
a=\frac{1}{4\omega^2}\left ( 4\omega_k^2-6\lambda A_0^2-\gamma^2 \right ),
~~~~
q=\frac{3\lambda A_0^2}{4\omega^2}.
\label{q}
\end{equation}
As is well-known \cite{ARN}, the Mathieu equation exhibits
parametric resonances when $\sqrt{a} \simeq i$, where $i=1,2,3,...$.
The width of the resonance regions depends on the ratio $q/a$ (see,
e.g. Ref. \cite{stegun}). In the framework of Eq.(\ref{MATHIEU1}) the
extent of the primary resonance $a\simeq 1$ can easily be
 estimated \cite{stegun} to be $(a-1)^2<q^2$. However, in the presence of the damping
$\gamma$ the resonance condition for Eq.(\ref{HILL1}) becomes
\begin{equation}
q^2> \frac{\gamma^2}{\omega^2}+(a-1)^2.
\label{basicreso}
\end{equation}
With $a$ and $q$ defined in Eqs. (\ref{q}), given
$\lambda,\gamma,\omega,\omega_0$, and $\varepsilon$, this translates into
an instability band of certain wavenumbers $k$.

The above analysis is easily extended to the case of two spatial
dimensions \cite{2Dana}, the only required change being that the
dispersion relation now is
$\omega^2_{\vec{k}}=\omega_0^2+4\sin^2(k_x/2)+4\sin^2(k_y/2)$, where
the wavevector is $\vec{k}=(k_x,k_y)$. The instability in this case
appears on an annulus in the wavevector plane, with a radius given by
$a=1$ (see, Eq. (\ref{q})) and a width determined by
Eq. (\ref{basicreso})

Numerical simulations allow us to follow the full
nonlinear development and saturation initiated by the linear instability itself. In
particular, in regions of parameter space we obtain the spontaneous
formation of patterns of distinct spatial geometry. 
Although we have studied this phenomenon in one as well as in two
dimensions, in the present communication we focus on the 
two-dimensional system, where the pattern formation is particularly rich.
 
Although the dynamics show different features according to the specific
region of parameter space, it is possible to trace a typical 
behavior as 
follows: Initializing the system in the spatially homogeneous state 
described above with a small amount of randomness added, 
the instability sets in after a certain number of cycles of the
ac-drive depending on the strength of the parametric resonance, 
{\em i.e. }on the value of $\sqrt{a} \simeq 1,2,3,...$.
Thereafter the system usually evolves through a sequence of 
different patterns (rhombi, stripes, etc.),
composed of localized regions of high amplitude oscillations,
before reaching a final configuration that may or may not result in a 
structure of definite symmetry.

\begin{figure}[h]
\vspace*{300pt}
\hspace{20pt}
\includegraphics{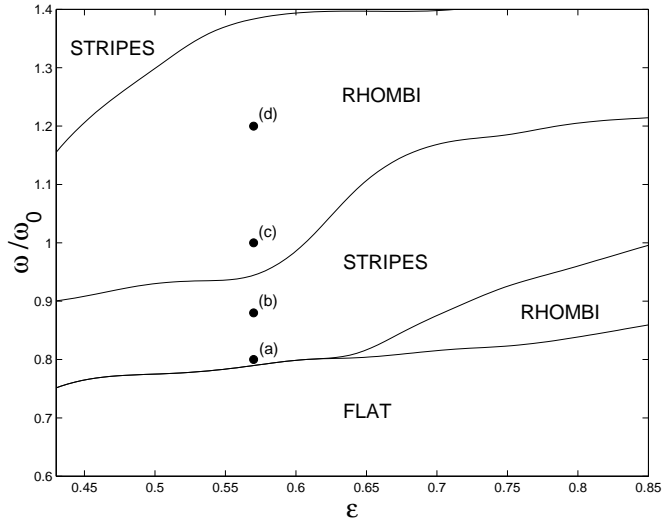}
\vspace*{-100pt}
\caption{Stability diagram showing the stability boundaries between the
  possible patterns in the ($\varepsilon$,$\omega$)-plane. 
  Parameters are $\omega_0=0.75$, $\lambda=-1$ (hard
  potential), $\gamma=0.05\omega_0$. The four points refer to the
  specific spatial structures shown in Fig. \ref{fig2}.}
\label{fig1}
\end{figure}

Due to the sensitive response to very small
changes in the parameters, determining stability regions for the
different pattern geometries generally is a difficult task. 
However, in the case of a hard potential ($\lambda=-1$), Fig.\ref{fig1}
shows a limited area of 
($\varepsilon$,$\omega$)-space in which distinct spatial patterns emerge and
remain stable.

\begin{figure}[h]
\vspace*{410pt}
\hspace{-40pt}
\includegraphics{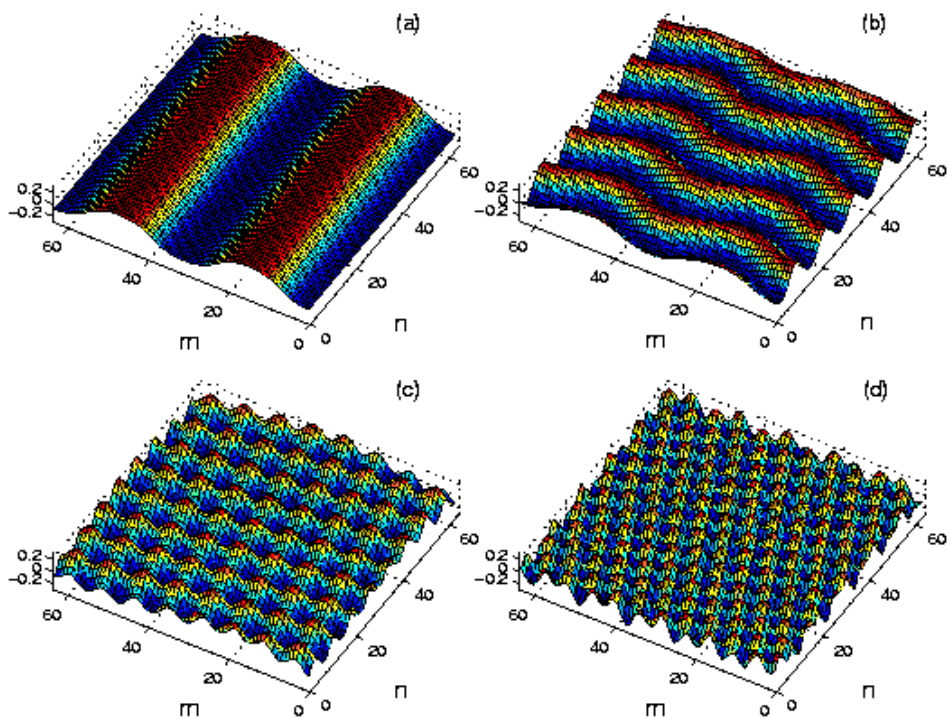}
\vspace*{-130pt}
\caption{(Color) Snapshots of spatial patterns corresponding to the four points marked in
  Fig. \ref{fig1} for $\varepsilon=0.57$: (a) straight stripes ($\omega=0.8\omega_0$),
  (b) modulated stripes ($\omega=0.88\omega_0$), (c) rhombi ($\omega=1.0
  \omega_0$), (d) localized rhombi ($\omega=1.2\omega_0$). $x_{n,m}$
  is plotted along the vertical axes.}
\label{fig2}
\end{figure}

This diagram is constructed following the full dynamics of the system
for thousands of cycles. As in the case of Ref. \cite{austin}, hysteresis is likely to occur
but we do not pursue this further here. 
Figure \ref{fig2} shows representative snapshots of 
the spontaneously emerging patterns corresponding to the points marked in Fig. \ref{fig1}.

The patterns consist of localized regions of high amplitude 
oscillations (i.e. intrinsic localized modes (ILM's)) residing on a background that oscillates 
at the frequency $\omega$ of the ac-drive.
In all the cases considered we have observed the natural result that
patterns are only energetically sustained when the ILM,
$\omega_{ILM}$, and driving, $\omega$, frequencies are commensurate, 
{\em i.e. }$\omega_{ILM}=n\omega$, where $n$ is an integer. For the 
patterns displayed in Fig. \ref{fig2}, $n=2$.
Furthermore, the motion of the ILM's is exactly out of phase, 
meaning that at times where the background oscillation 
reaches its maximal excursion the ILM's attain their minimal 
amplitude such that at these points the state is completely
homogeneous.

At a fixed driving $\varepsilon$, for increasing values of the frequency $\omega$, as in 
Fig. \ref{fig1}, we observe the following behavior: Due to the presence of the damping 
$\gamma$, at sufficiently small $\omega$ the spatially homogeneous solution is 
stable towards all possible spatial modulations such that the {\em flat} state is sustained. However, for
values near point (a) the system becomes unstable with respect to certain 
spatial modulations and spatial
patterns in the form of broad stripes emerge (Fig. \ref{fig2}(a)). Increasing the 
frequency, these stripes become thinner and denser and begin to show an increasingly 
clear modulation (Fig. \ref{fig2}(b)). The characteristic length 
scale of these patterns is set by the size of the unstable $\vec{k}$-vector according to the above analysis.
The nonlinear character of the system results in a transition 
towards a more isotropic geometry (rhombic) as the driving frequency 
is increased further (Fig. \ref{fig2}(c)). As in the case of the stripes, stronger localization of the ILM's arranged in
the rhombic pattern (Fig. \ref{fig2}(d)) is observed for even larger driving frequencies. The angle between the 
sides of the rhombus' unit cell varies but for the hard potential it is always 
close to $\pi/2$. For values of $\varepsilon$ larger than
those displayed in Fig. \ref{fig1}, the final mesoscopic patterns of the system 
dynamics are spatially disordered much like the phenomena observed in
granular media \cite{austin}. 
It is important to realize that the 
length of the unstable wavevector determines the length-scales of the final patterns, while 
the symmetry of the patterns is determined by the nonlinearity of the system. 

For the soft potential ($\lambda=1$) the variation of the amplitude $\varepsilon$, and
the frequency $\omega$, of the ac-drive is particularly problematic since
the dynamics in this case can lead to the development of catastrophic instabilities as 
one or several oscillators overcome the finite barrier in the quartic 
potential. In all the cases we have been able to simulate, the early stage time
evolution of the system is characterized by the formation of ILM's 
regularly arranged in a square pattern. This spatial configuration,
which is sustained for up to hundreds of cycles, seems always to
suffer from a weak instability and eventually deforms into a rhombic
pattern. In contrast to the case of the hard potential, with the soft
potential the angle of the rhombus unit cell is always close to $2\pi/3$ (so
almost hexagonal). This difference can be understood by exploiting
the analogy between the changes in the steady states of  dissipative systems
and phase transitions in systems at thermodynamic equilibrium\cite{boris}.

\section{Dry friction}

Spurred on by recent technological advances that have given  new 
insight into frictional processes at the atomistic level \cite{Jref1}, 
there have recently
been many attempts to model atomic-scale friction using low-dimensional
nonlinear models \cite{Jref2}. 
Many of these have been variations on driven one- or
two-dimensional 
Frenkel-Kontorova (FK) type models, in which a chain or layer of interacting 
particles is subject to a periodic substrate potential and a driving force.
Here (see also Ref. \cite{Jref3}) we describe one model which attempts to go
one step further in coupling the nonlinear atomic dynamics which occur at
the sliding interface with the essentially linear behavior deep in the 
bulk of the materials.

\begin{figure}[h]
\vspace*{-5pt}
\hspace{0pt}
\includegraphics{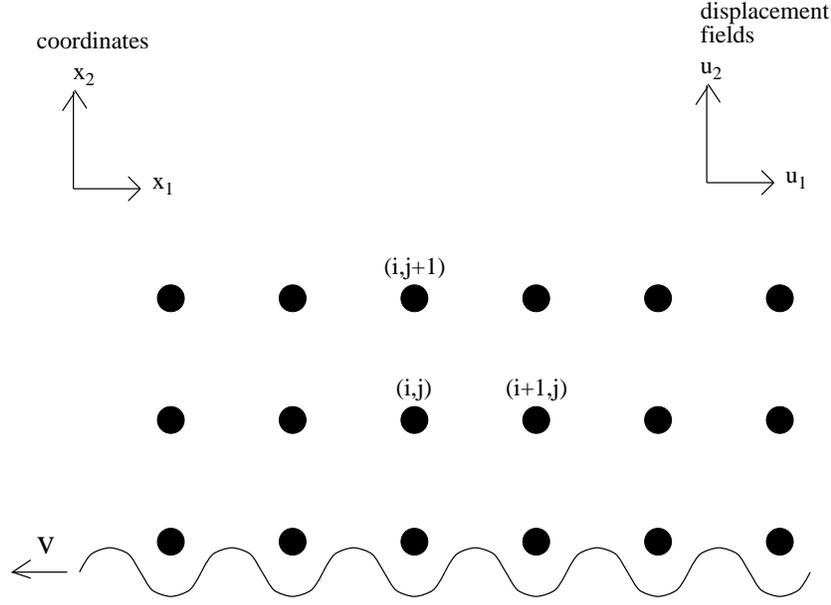}
\vspace*{230pt}
\caption{Geometry of the ``isotropic elastic block'' model of sliding 
friction described in the text.}
\label{Jfig1}
\end{figure}

In the model the upper workpiece is represented by a one-dimensional
chain of atoms at the interface coupled to an isotropic elastic
medium. The lower workpiece is replaced by a sinusoidal substrate
potential. Relative motion between the two workpieces is produced by
translating the substrate potential at a fixed velocity, $v$.
The two-dimensional (2D) medium is represented by a 
2D array of 2D displacements
$u_{1}(i,j)$ and $u_{2}(i,j)$, for $i$ and $j$ spanning the medium.
  The geometry of the model is shown in Fig.~\ref{Jfig1}.

The equations of motion of the model can be written as
\begin{eqnarray}
\ddot{u}_1(i,j) &=& ( \lambda + \mu ) ( u_{1,11}(i,j)+u_{2,21}(i,j) )
 \nonumber \\
&+& \mu ( u_{1,11}(i,j) + u_{1,22}(i,j) )
   + \frac{\delta_{1j}}{2 \pi} \sin ( 2 \pi 
( u_1(i,j) +  vt )) \nonumber \\
\ddot{u}_2(i,j) &=& ( \lambda + \mu ) ( u_{2,22}(i,j)+u_{1,12}(i,j) )
\nonumber \\
&+& \mu ( u_{2,11}(i,j) + u_{2,22}(i,j) ) ,  
\end{eqnarray}
where $u_{l,mn}(i,j)$ is the discretized second derivative of the displacement
component $u_l$ at the site $(i,j)$ in the directions $m$ and $n$. 
The Lam\'e constants, $\lambda$ and $\mu$ were taken to be $1/3$, so that
the longitudinal sound speed, $c_l=1$.
Notice that only the first layer of the block interacts directly with the 
substrate
potential, confining the nonlinearity  to  the interface.
These equations can be solved numerically to yield the long-time
steady state behavior of the system at any given $v$.

\begin{figure}[h]
\vspace*{330pt}
\hspace{-30pt}
\includegraphics{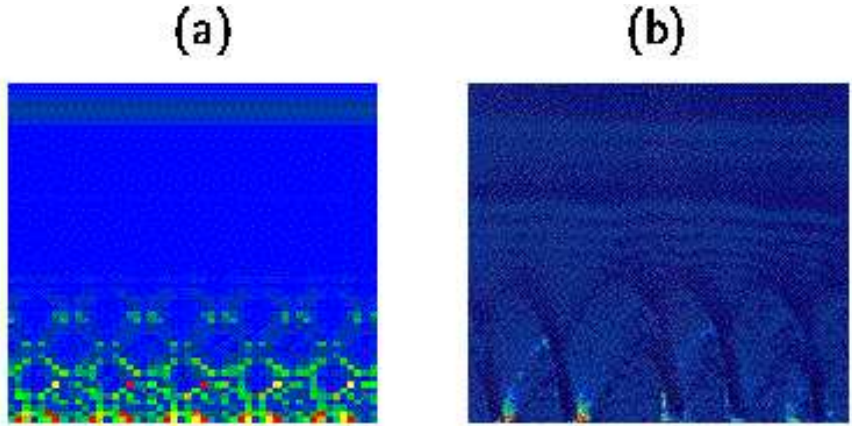}
\vspace*{-150pt}
\caption{(Color) (a) Snapshot of the elastic energy of the model; (b) Snapshot of 
the potential energy of the 2D molecular dynamics simulations. 
(Both are color coded from
blue at low values to red at high values.)}
\label{Jfig2}
\end{figure}

For small enough
velocities, i.e. $v<v_c \approx 0.2$, the block sticks to the substrate 
and will be dragged along
with the same average velocity.
For higher
velocities, $v>v_c$, one observes that there is an initial jerk, the size of
which depends on how close the block is to sticking to the substrate. 
After this
jerk, the block oscillates around a fixed position. For perfectly uniform
initial conditions (i.e. all $u_{1}=u_2=0$) and no noise this persists as 
the steady state. However, when one allows for
 a small random perturbation 
on the uniform initial conditions, 
one observes that during the 
oscillation a certain wavelength modulation in the $x_1$ direction is selected.
This modulation grows until saturation. Then one observes that the block 
assumes a small non-zero velocity to reach the steady state.
 This time development can easily
 be followed by looking at a snapshot of the elastic energy.
As can be seen in the Fig.~\ref{Jfig2}(a), the leading wavefronts are practically flat,
then the modulation develops until it saturates. One then observes that 
 a periodic pattern develops in the elastic medium. 
Similar patterning has been observed in one-dimensional 
FK and sine-Gordon systems
\cite{Jref4}.

The selection of a particular wavelength parallel to the interface,
 $k_{x_1}$, is
apparent in Fig.~\ref{Jfig2}(a). This wavelength is chosen by the system
matching the driving frequency of the moving substrate potential as closely
as possible
to the dispersion relation of the longitudinal waves of the elastic medium,
$\omega = c_l k_{x_1}$. Perfect agreement is achieved in this process
by the block adopting a small non-zero velocity, $v_{block}$  so  that
the driving frequency is tuned to
$ \pi v_{slip}= \pi (v-| \langle v_{block} \rangle |)= c_l k_{x_1}$.
(Angular brackets represent a time-averaged quantity.)

\begin{figure}[h]
\vspace*{-50pt}
\hspace{-25pt}
\includegraphics{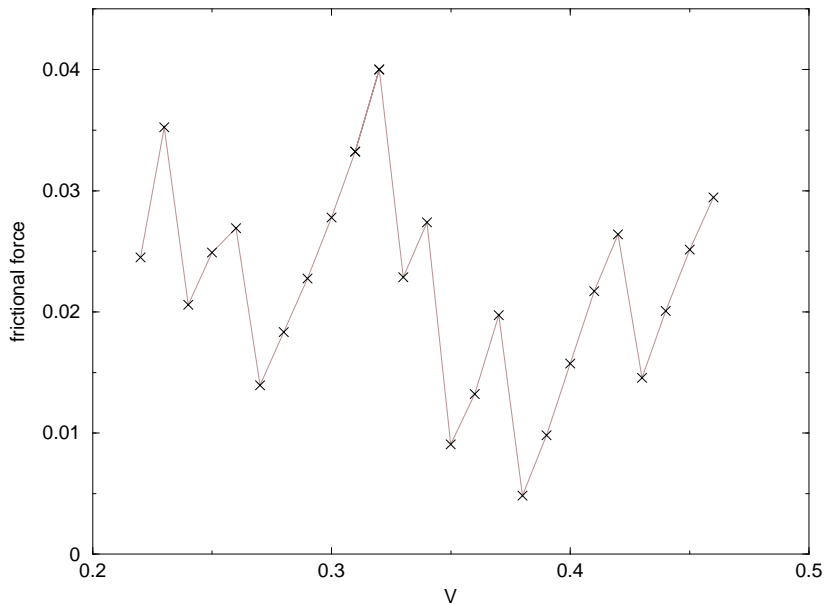}
\vspace*{290pt}
\caption{Plot of the frictional force as a function of substrate velocity,
$v$.}
\label{Jfig3}
\end{figure}

This effect of locking to the wavelength most closely matched to the
driving frequency can be seen in the frictional behavior of the system.
Fig.~\ref{Jfig3} shows the frictional force exerted by the substrate on the
elastic block as a function of $v$. One sees regular
portions of the curve, where one particular wavelength is selected,
separated by jumps where the system changes from one modulation
wavelength to the next. So, the spatial patterning is
reflected in the mechanical properties of the system.

Large-scale molecular dynamics (MD) simulations have been performed
of the sliding of two 2D copper workpieces incorporating realistic embed-ded-atom 
potentials. An early-time
 snapshot of the potential
energies of the atoms in the simulations is shown in Fig.~\ref{Jfig2}(b),
alongside a corresponding snapshot of the elastic energies for the model.
It is evident that the simple model successfully reproduces the
qualitative features of the early-time behavior of the MD simulations.
In addition, the elastic block model is able to elucidate the {\em physical
mechanism} behind the patterning observed in the large-scale MD simulations.

\section{Patterns arising from elastic compatibility}
Martensitic structural transitions \cite{R1}, especially those with
unit cells related by continuous deformations (solid-solid phase transitions), exhibit a rich variety of
temperature/stress induced microtextures \cite{R2,R3,R4}.  For
example, alloys such as FePd, and NiTi show the shape memory effect \cite{R5},
transforming on cooling from a higher symmetry phase (`austenite'), through
nanometer-scale `tweed' textures; to equal-width mesoscale `twins'
below the (`martensite') transition temperature $T_0$.

To understand martensitic transitions and textures in a Ginzburg-Landau (GL)
approach based on the natural order parameter (OP), we need a treatment
in terms of the physical strain variables alone.
The OP is one or more components of a strain tensor $\varepsilon_{\mu\nu}({\mu,\nu=x,y,z})$;
not a true scalar. For simplicity, we consider here a 2D square-to-rectangle
transformation \cite{R6,R7,R10,R11}, with a rectangular or deviatoric strain as
the OP. The non-OP strain components are implicit functions of the OP, through a
compatibility (differential) equation \cite{R12}.
Thus the apparently innocuous GL terms, harmonic in the non-OP
components, are crucial: they generate two effective long-range anisotropic OP
potentials: from the bulk and from the `habit-plane' interfaces. The combined action
of these {\it two} compatibility potentials plays a decisive role in the energetic competition
between various OP textures, e.g. for temperatures $T<T_0$, oriented,
{\it equal-width}
twins are favored, emerging into the bulk from the habit plane.
Our GL model includes alloy composition fluctuations as local internal
micro-stresses, that induce tweed textures at $T>T_0$. We show that simple and local
external stresses can generate complex and extended, multiscale structures throughout
the system, that could play a role in shape memory. Thus we find that
bulk/interface strain potentials induced by
compatibility
enable a unified description of spontaneously formed, and stress-induced
martensitic textures, in a GL model in terms of the OP strain alone.
Our GL model in 2D consists of (i) a triple-well potential
$F_0$, as is usual for first order transitions, in the
deviatoric strain OP
[$\epsilon=(1/\sqrt{2})(\varepsilon_{xx}-\varepsilon_{yy})$],
(ii) harmonic (linear) elastic
energy cost $F_{cs}$ due to the compressional
[$e_1=(1/\sqrt{2})(\varepsilon_{xx}+\varepsilon_{yy})$]
and shear ($e_2=\varepsilon_{xy}$) strain that implicitly depend on the
OP through compatibility; (iii) coupling of strain(s) to an external
or internal (defect, dislocation) stress; (iv) second
and fourth order strain gradient terms $F_{grad}$ that give rise to
multiscale competition;
and (v) symmetry-allowed couplings
$F_{compos}$ of the (scalar) compositional fluctuations ${\tilde\eta}(r)$
to the OP and its derivatives.

The (dimensionless) elastic energy in 2D is given with,
$\epsilon = \epsilon(r)$, by:
\begin{equation}
F=F_0(\epsilon)+F_{grad}(\nabla\epsilon)+F_{cs}(e_1,e_2)+
F_{compos}(\epsilon), 
\end{equation}
where the detailed expressions can be found in Ref. \cite{RAPID}.
The dynamics of the continuous OP is assumed to be of the time dependent GL
or relaxational type,
\begin{equation}
\dot{\epsilon}(r)=-\frac{\partial F(\{\epsilon(r),e_1(\epsilon(r)),
e_2(\epsilon(r))\})}{\partial \epsilon(r)},
\end{equation}
where time $t$ is scaled with a characteristic relaxation rate.  The
compres-sion-shear (CS) strains $e_1(r), e_2(r)$ are written in terms of
the order parameter $\epsilon(r)$ by solving the elastic compatibility
(differential) equation.
The analysis with
six strain
tensor components in 3D can be carried out \cite{R15}, but for simplicity, we confine our
discussion to 2D with a
compatibility constraint, satisfied
at all times \cite{R7,R16}: 
\begin{equation}
\nabla^2e_1(r)-\sqrt{8}\nabla_x\nabla_ye_2(r)=(\nabla_x^2-\nabla_y^2)
\epsilon(r). 
\end{equation}
For (Fourier-expandable) strains $\epsilon(r)$ in the bulk, one obtains
 via the Lagrangian multiplier formalism \cite{R7} that $e_1(\vec k), e_2(\vec k)$
are proportional
to $\epsilon(\vec k)$, with $\vec k$ dependent coefficients.
An OP strain-strain  potential
$F{(\epsilon)}_{cs} = F^{(bulk)}_{cs}+F^{(surface)}_{cs}$ replaces $F_{cs}(e_1,e_2)$, where
$F_{cs}^{(bulk)}=\sum_k U^{(bulk)}(\vec k)|\epsilon(\vec k)|^2$ and in $\vec{k}$-space
\begin{equation}
U^{(bulk)}(\vec k)= \frac{\frac{2A_1}{A_2}[\frac{(k_x^2-k_y^2)}{k^2}]^2}{1+
\frac{8A_1}{A_2} \frac{k_x^2k_y^2}{k^4}}.
\end{equation}
Here $A_1$ and $A_2$ are bulk and shear modulus, respectively.
A similar expression can be derived for the surface contribution $F^{(surface)}_{cs}$.

In the numerical simulations reported below we take random initial conditions, a 
$256\times 256$ lattice, and periodic boundary conditions in both $x$ and 
$y$ directions to obtain relaxed textures $\epsilon(r)$. 
The red/blue/green color shades represent $\epsilon$ positive/negative/zero 
OP strain values.

\begin{figure}[h]
\vspace*{345pt}
\hspace{20pt}
\includegraphics{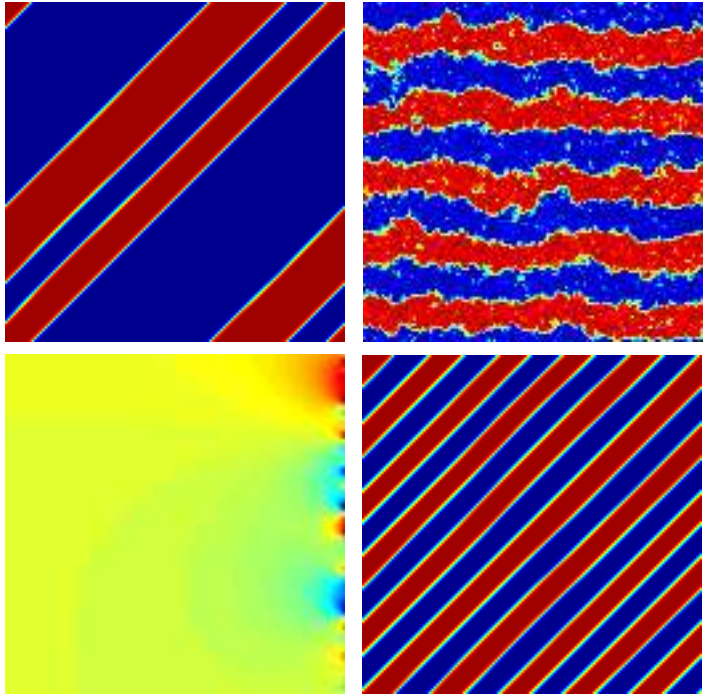}
\vspace*{-70pt}
\caption{(Color) Clockwise from upper left: (a) Oriented ${\pi}/{4}$ parallel
structure obtained from bulk compatibility only.
(b) Unoriented `twins' obtained from surface compatibility only. (c)
Oriented twins with both bulk and surface compatibility present. (d)
Corresponding elastic fringing field in the austenite. }
\label{MA1}
\end{figure}

Figure \ref{MA1} shows formation of low temperature $T<T_0$ parallel-domain 
structures with only bulk compatibility included. Note that the structures 
have the proper 45$^{\circ}$ orientation (as in \cite{R7}) but
are not true twins:
there is no 
well 
defined width scale. In Fig.~\ref{MA1}b we 
depict
horizontal
twin-like structures with only the surface compatibility 
term included.  Now there is a dominant length scale (twin width) 
in the system but the `twins' have rough interfaces and are not properly 
oriented. When {\it both} bulk and surface compatibility terms are included 
we obtain equally spaced, parallel, and properly oriented twins.  The corresponding 
derived
(deviatoric) OP strain field
$\tilde {\epsilon}(r)$
in the austenite 
\cite{R6} falls off as `decaying twins' away from the habit plane, as  
depicted in Fig.~\ref{MA1}(d). 

\begin{figure}[h]
\vspace*{385pt}
\hspace{20pt}
\includegraphics{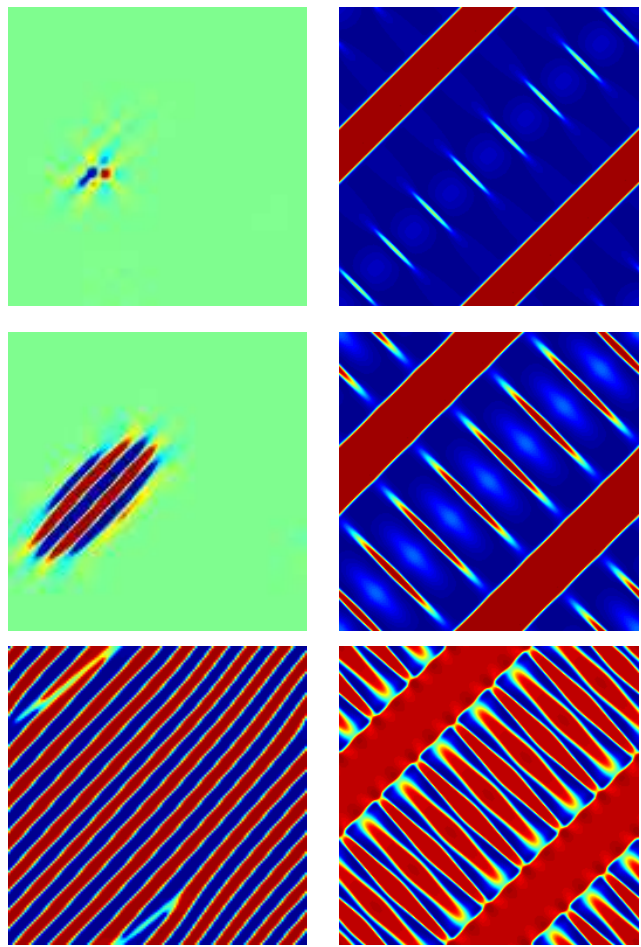}
\vspace*{-10pt}
\caption{(Color) Effect of a temperature quench and stress: (a) Left column, top down, a time
sequence showing how a `temperature quench' of austenite
results in nucleation and growth of twinned martensite,
(b) Right column, top down, a time sequence under
external uniform stress (seeded) twins, inducing
hierarchy and branching of twins.}
\label{MA2}
\end{figure}

Figure~\ref{MA2}a shows how $T >> T_0$ austenite, evolves to twinned martensite, under
a temperature quench to $ T < T_0$. There is a growth process resulting 
in twinned martensite. Note the lenticular
twin shapes, surrounding elastic field and the ``skew-varicose" interface
instability just before the twinning is complete.  The latter instability
is observed in many other contexts such as granular media and convective rolls
in fluids \cite{R18}. The total compressional/shear
strains, derivable from the bulk OP strain through compatibility, as above,
 are not
shown here. For $T < T_0$, $e_1$, $e_2$ are expelled  from the bulk to the
interface when
equilibrium 
OP twins with $\pi/4$ orientation finally emerge.
 By contrast, the $T>T_0$ OP tweed shows $e_1$, $e_2$ strains in the bulk, localized at the
domain-wall crossing points. This is reminiscent of type-II superconductors,
with $T_0 \sim T_{c1}$, and with twin (tweed) textures acting like
Meissner (vortex) phases, expelling (allowing localized) shear strains 
that are analogous to transverse vector potentials.

Fig.~\ref{MA2}b shows how (suitably seeded) twins can evolve under uniform external stress. The growth of (at
least two generations of) hierarchical twinning  is shown .  
A new generation of twins nucleates
when the elastic strain between two preexisting twins exceeds a critical value.
Branched twin-like structures may be seen where the fine twins meet the 
coarser ones.

\section{Conclusions}
By way of three quite different examples we have illustrated some of the salient features
controlling mesoscopic 
patterns: Forced versus unforced environments; the effects of discrete lattice scales; the competitions between
anisotropic short and long-range interactions; and the manifestation of constraints as source of long-range interactions.
The remaining challenge, once one fully understands how to model systems at mesoscopic scales, is the length scale
bridging: (i) To feed information 
in an appropriate and feasible way from experiments and microscopic models such as molecular dynamics simulations, 
electronic structure calculations etc. into the mesoscopic models. (ii) Use the information obtained 
from mesoscopic models as building components in understanding macroscopic behavior in the context of, for example, 
mechanical finite element models or the materials consitutive response.

\begin{acknowledgments} 
We gratefully acknowledge fruitful discussions with V. Bortolani, J. Hammerberg, B.L. Holian, B.A. Malomed,
R. Mikulla, and S.R. Shenoy. 
Work at Los Alamos National Laboratory is performed under the auspices of the US DOE.
\end{acknowledgments}

\bibliographystyle{apalike}

\begin{chapthebibliography}{1}

\bibitem{R1} Z. Nishiyama, {\it Martensitic Transformations} (Academic,
New York, 1978).

\bibitem{Andrea} A. Vanossi, K.~{\O}. Rasmussen, A.~R. Bishop,
  B.~ A. Malomed, and V. Bortolani, Phys. Rev. E, {\bf 62}, 7353 (2000).


\bibitem{2Dana} The two-dimensional version of Eq.(\ref{eqs}) is
  $\ddot x_{n,m}+\gamma \dot x_{n,m} +\omega_0^2 x_{n,m}=
  x_{n+1,m}-4x_{n,m}+x_{n-1,m}+x_{n,m-1}+x_{n,m+1} +\lambda x^3_{n,m}
  +\epsilon \cos\omega t$, where the parameters have the
  interpretation given in connection with Eq. (\ref{eqs}).


\bibitem{ARN} V.~I. Arnold, {\em Mathematical Methods of Classical
    Mechanics}, (Springer, New York, 1997).

\bibitem{stegun} M. Abramowitz and I. ~A. Stegun, {\em Handbook of
    mathematical functions with formulas, graphs, and mathematical
    tables}, (U.S. Govt. Print. Off., Washington, D.C., 1964).

\bibitem{austin} F. Melo, P. Umbanhowar, and H.~L. Swinney, Phys.
  Rev. Lett. {\bf 72}, 172 (1994); {\bf 75}, 3838 (1995); P.
  Umbanhowar, F. Melo and H.~L. Swinney, Nature, {\bf 382}, 793
  (1996); C. Bizon, M.~D. Shattuck, J.~B. Swift, W.~D. McCormick, H.~L. Swinney, Phys. Rev. Lett. {\bf 80}, 57 (1998).

\bibitem{boris} B.~A. Malomed, A.~A. Nepomnyashcii, and T.~I. Tribekskii,
 Zh. Eksp. Teor. Fiz. {\bf 96}, 684 (1989). [Sov. Phys. JETP {\bf 69}, 388 (1989)].

\bibitem{Jref1} B. Brushan, J.~N. Israelachvili, and U. Landman, Nature 
(London) {\bf 374}, 607 (1995).

\bibitem{Jref2} B.~N.~J. Persson, {\em The Physics of Sliding Friction},
(Springer, Heidelberg, 1998); M.~O. Robbins and M.~H. M\"user, in
{\em Handbook of Modern Tribology} (CRC Press, 2000).

\bibitem{Jref3} J. R\"oder, A.~R. Bishop, B.~L. Holian, J.~E. Hammerberg,
and R.~P. Mikulla, Physica D {\bf 142}, 306 (2000).

\bibitem{Jref4} J. R\"oder, J.~E. Hammerberg, B.~L. Holian, and A.~R. Bishop,
Phys. Rev. B {\bf 57}, 2759 (1998); J.~C. Ariyasu and A.~R. Bishop, Phys.
Rev. B {\bf 35}, 3207 (1987).



\bibitem{R2}
L. E. Tanner, A. R. Pelton, and R. Gronsky,
J. Phys. (Paris) {\bf 43},  C4-169 (1982).

\bibitem{R3}
R.~Oshima,  M.~Sugiyama, and F.E.~Fujita,
Metall. Trans. A {\bf 19}, 803 (1988).

\bibitem{R4}
M.~Sugiyama, Ph.~D. Thesis, Osaka University, 1985.

\bibitem{R5}
{\it Shape Memory Effect in Alloys}, J. Perkins, ed. (Plenum, New York,
1975); {\em Shape Memory Materials}, eds. K. Otsuka, and C.~M. Wayman, (Cambridge University Press, Cambridge, 1998).

\bibitem{R6}
G. R.~Barsch, B.~Horovitz, and J. A.~Krumhansl,
Phys. Rev. Lett. {\bf 59}, 1251 (1987); B.~Horovitz, G. R.~Barsch, and J. A.~Krumhansl,
Phys. Rev. B {\bf 43}, 1021 (1991).

\bibitem{R7}
S.~Kartha, T.~Kast\`an, J.A.~Krumhansl, and J.P.~Sethna,
Phys. Rev. Lett. {\bf 67}, 3630 (1991); J. P.~Sethna et al.,
S.~Kartha, T.~Kast\`an, and J.A.  Krumhansl,
Phys. Scripta {\bf T42}, 214 (1992); S.~Kartha, J.A.~Krumhansl, J.P.~Sethna, and L. K. Wickham,
Phys. Rev. B {\bf 52}, 803 (1995).

\bibitem{R10} A. E. Jacobs, Phys. Rev. B {\bf 31}, 5985 (1985); {\bf 46}, 8080 (1992);
{\bf 52}, 6327 (1995).

\bibitem{R11}
A. Saxena, S.~R. Shenoy, A.~R. Bishop, Y.~Wu, and T. Lookman,
Physica A {\bf 239}, 18 (1997).

\bibitem{R12}
D. S. Chandrasekharaiah and L. Debnath, {\it Continuum Mechanics}
(Academic Press, San Diego, 1994), p 218.

\bibitem{RAPID} S.~R. Shenoy, T. Lookman, A. Saxena, and A.~R. Bishop, Phys. Rev. B, {\bf 60}, R12537 (1999).

\bibitem{R15}  K.~{\O}. Rasmussen, T. Lookman, A. Saxena, A.~R. Bishop, and R.~C. Albers,
    http://xxx.lanl.gov/abs/cond-mat/0001410.

\bibitem{R16}
W.~C. Kerr, M.~G. Killough, A. Saxena, P.~J. Swart, A.~R. Bishop, Phase Trans. {\bf 69}, 247 (1999).
\bibitem{R18}
J.~R. deBruyn, C. Bizon, M.~D. Shattuck, D. Goldman, J.~B. Swift, H.~L. Swinney,
Phys. Rev. Lett. {\bf 81}, 1421 (1998).
\end{chapthebibliography}

\notes

\end{document}